\documentclass[pra,twocolumn,showpacs,groupedaddress,superscriptaddress,aps,10pt]{revtex4-1}
\usepackage{bm,graphicx,amsmath}
\usepackage{placeins}
\usepackage{amssymb}
\usepackage{amsmath}
\usepackage{epsfig}
\usepackage{amssymb}
\usepackage{color}
\usepackage[colorlinks,linkcolor=blue,citecolor=blue]{hyperref}
\usepackage{subfigure}

\begin{document}

\title{Multiple rings formation in cascaded conical refraction}
\date{\today}

\author{A. Turpin}
\affiliation{Departament de F\'isica, Universitat Aut\`onoma de Barcelona, Bellaterra, E-08193, Spain}
\author{Yu. V. Loiko}
\affiliation{Departament de F\'isica, Universitat Aut\`onoma de Barcelona, Bellaterra, E-08193, Spain}
\author{T. K. Kalkandjiev}
\affiliation{Departament de F\'isica, Universitat Aut\`onoma de Barcelona, Bellaterra, E-08193, Spain}
\affiliation{Conerefringent Optics SL, Avda. Cubelles 28, Vilanova i la Geltr\'u, E-08800, Spain}
\author{J. Mompart}
\affiliation{Departament de F\'isica, Universitat Aut\`onoma de Barcelona, Bellaterra, E-08193, Spain}

\begin{abstract} 
When a light beam passes through a cascade of biaxial crystals with aligned optic axes, the resulting transverse intensity pattern consists of multiple concentric rings. We provide a simple formulation for the pattern formation for both circularly and linearly polarized input beams, that could be applied for a cascade of an arbitrary number of biaxial crystals. We have experimentally investigated multiple ring formation with up to three cascade biaxial crystals, showing that the theoretical formulation is in full agreement with the experimental results. \\
\textbf{ocis}: 160.1190, 260.1180, 260.1440.
\end{abstract}

\date{\today}
\maketitle

\begin{figure}[]
\centerline{\includegraphics[width=0.70 \columnwidth]{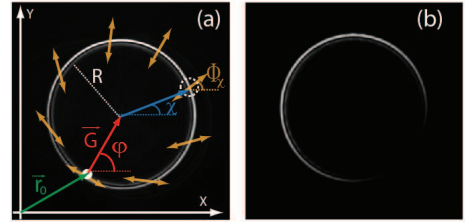}}
\caption{(color online) (a) CR ring at the focal plane for a circularly polarized input beam. Orange double arrows show the polarization distribution at each point of the CR ring. See text for parameters' definition.
(b) CR pattern for a vertically linearly polarized input beam.}
\label{fig1}
\end{figure}

In conical refraction (CR), when a circularly polarized focused light beam passes along one of the optic axis of a biaxial crystal (BC), it is transformed into a light ring at the focal plane of the system \cite{kal2008,Berry2}. This light ring splits into a pair of concentric bright rings separated by a dark (Poggendorff) ring under conditions of $R \gg w_0$, where $R$ and $w_0$ are the radius of the CR ring and the waist radius of the focused input beam, respectively. One interesting feature of the CR ring is that it is laterally shifted with respect to the input beam. Both the direction and magnitude of this lateral shift can be represented by the characteristic vector of the crystal, $\vec{G} \equiv R \left [ \cos(\varphi),\sin(\varphi) \right]$, that belongs to the plane of the crystal optic axes, where $\varphi$ is the orientation of the plane of optic axes \cite{kal2008}, see Fig.~\ref{fig1}(a). 
The polarization distribution along the ring depends only on the orientation of $\vec{G}$. Each point of the light ring is linearly polarized with the polarization plane rotating so that every pair of diagonally opposite points of the ring have orthogonal polarizations. As a consequence, for randomly and circularly polarized (RP/CP) input Gaussian beams, light's intensity distribution along the ring is uniform. In contrast, if the input beam is linearly polarized (LP), the CR ring possesses a point of null intensity whose polarization is perpendicular to the input beam's one, see Fig.~\ref{fig1}(b).

Recently, multiple concentric rings have been experimentally reported in a cascade of two \cite{kal2008,phe2012,CRlaser} and up to four \cite{abd2011, abd2012} BCs, with interesting applications in lasing \cite{CRlaser}, particle trapping \cite{odw2010,Vortex} and free space optical communications \cite{CRFSOC_OL}. The diffracting wave theory of CR \cite{Belskii1, Berry2} has been extended to the case of beam propagation through a cascade of up to $N$ BCs \cite{phe2012,Berry3}, providing an accurate description of the phenomenon in terms of Bessel functions. Even so, a simpler approach should be developed to offer a practical guideline of the final pattern for CR in cascaded BCs, including the case of linearly polarized input beams that has not been reported yet. This is the main goal of this article.

Here we present experiments and a simple theoretical insight into the problem of multiple ring pattern formation in a cascade of BCs. Our approach comes from the transformation rules based on wave-vector and polarization dependence of CR reported recently \cite{ebs2013} and on the experiments with spatially filtered beams obtained by properly selecting an angular part of the CR ring with a pinhole placed at the focal plane \cite{Loiko2011} of the first BC. 

\begin{figure}[]
\centerline{\includegraphics[width=1 \columnwidth]{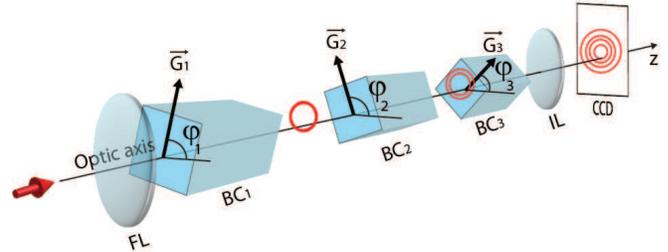}}
\caption{(color online) Experimental set-up. An input beam propagating along the optic axis of a cascade of biaxial crystals (BC$_i$) (with orientation $\varphi_{i}$) is focused with a lens (FL) and transformed by the CR phenomenon in the cascade. 
An imaging lens (IL) projects the final pattern into the CCD camera.}
\label{fig2}
\end{figure}

The scheme of our experimental arrangement is shown in Fig.~\ref{fig2}. A circularly polarized Gaussian light beam is focused with a lens (FL), passing along the optic axis of a cascade of up to three biaxial crystals (BC$_i$) rotated by angles $\varphi_{i}$ around their aligned optic axes. Finally, an imaging lens (IL) projects the focal plane into the CCD camera.

\begin{figure}[!]
\centerline{\includegraphics[width=1 \columnwidth]{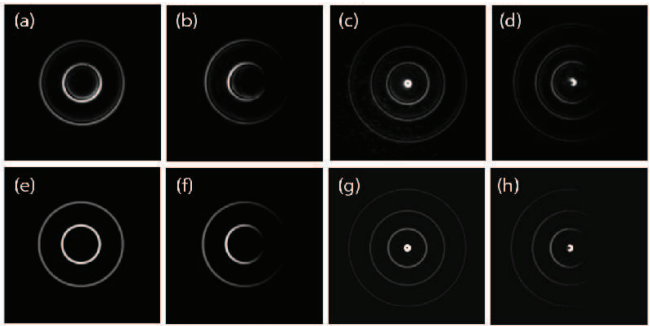}}
\caption{Transverse intensity patterns for a cascade of 2 (two concentric rings patterns) and 3 (four concentric rings patterns) BCs. Figures (a)-(d) are experimental data and (e)-(h) the corresponding theoretical simulations. First and third columns correspond to a circularly polarized input beam, while second and fourth columns correspond to a linearly polarized input beam with azimuth $\Phi_0 = 0^{\circ}$. Experimental parameters of the BCs: $L_1~=~27.31\,\rm{mm}$, $\varphi_{1}~=~0^{\circ}$; $L_2~=~10.66\,\rm{mm}$, $\varphi_{2}~=~90^{\circ}$; $L_3~=~18.29\,\rm{mm}$, $\varphi_{3}~=~180^{\circ}$. The focal length of the focusing lens (FL, see Fig.~\ref{fig2}) is $200\,\rm{mm}$.}
\label{fig5}
\end{figure}

Single annular pattern with fine Poggendorff splitting is observed when the first crystal is placed alone. When the second and third crystals are consecutively added, up to two and four concentric rings appear, as shown in Fig.~\ref{fig5}(a) and Fig.~\ref{fig5}(c), respectively. We introduce the transverse coordinates XY, see Fig.~\ref{fig1}(a). The BC$_1$ orientated at an angle $\varphi_{1}$ transforms the input Gaussian beam (with position vector $\vec{r}_0$) into the CR light ring, whose center position is given by 
\begin{equation}
\vec{r}_{1} = \vec{r}_0 + \vec{G}_1.
\label{eq1}
\end{equation}
From Eq.~(\ref{eq1}), the position of any point of the CR ring (represented by the polar angle $\chi$, see Fig.~\ref{fig1}(a)), is described by 
\begin{equation}
\vec{r}_{1}(\chi) = \vec{r}_0 + \vec{G}_1 + | \vec{G}_1 | \left [ \cos(\chi),\sin(\chi) \right ].
\label{eq2}
\end{equation}
The polarization azimuth, $\Phi_{\chi}$, at each point of the ring is related to its position along the CR ring through 
\begin{equation}
\Phi_{\chi} = \frac{\chi + \varphi_{1}}{2}.
\label{eq3}
\end{equation}

To investigate the origin of multiple rings generated by a cascade of BCs, part of the conically refracted beam is selected (filtered) by placing a pinhole at the CR ring after the first crystal, as schematically shown in Fig.~\ref{fig1}(a). Every point of the ring is a CR-filtered beam defined by the polar angle $\chi$ of the filtering (which also defines the plane of wave-vectors $\Phi_K=\chi$ \cite{ebs2013}) and by the plane of its electric field vector with polarization azimuth $\Phi_{\chi}$ given by Eq.~(\ref{eq3}).  
We observed that, as expected from Ref. \cite{Loiko2011}, CR-filtered beams passing through a BC do not produce full ring pattern, but refract (split) into two orthogonally linearly polarized beams. Their positions correspond to two diagonally opposite points of the otherwise expected CR ring for a Gaussian input beam. Their geometric center is 
\begin{equation}
\vec{r}_{2} = \vec{r}_{1} + \vec{G}_2 = \vec{r}_0 + \vec{G}_1 + \vec{G}_2, \label{eqaux}
\end{equation}
while their polar angles and transverse positions are defined by the filtering angle $\chi$ of the input beam 
\begin{eqnarray}
\chi_2^{(1)} = \chi,~ \chi_2^{(2)} = \chi+\pi; \label{eq4} \\
\vec{r}_2(\chi_{2}^{(1,2)}) = \vec{r}_{1}(\chi) + \vec{G}_2 \pm | \vec{G}_2 | \left [ \cos(\chi),\sin(\chi) \right ], \label{eq5}
\end{eqnarray}
where the superscript distinguishes the refracted (output) beams. Their polarization azimuths $\Phi_{\chi_{2}^{(1,2)}}$ are defined by Eq.~(\ref{eq3}) with $\chi$ being replaced by $\chi_{2}^{(1,2)}$ given in Eq.~(\ref{eq4}) and $\varphi_{1}$ being replaced by $\varphi_{2}$. By combining Eq.~(\ref{eq2}) with Eqs.~(\ref{eqaux}) and (\ref{eq5}), the latter reads
\begin{equation}
\vec{r}_2(\chi_2^{(1,2)}) = \vec{r}_{2} \pm R_{2}^{(1,2)} \left [ \cos(\chi),\sin(\chi) \right ], 
\label{eq6}
\end{equation}
where $R_{2}^{(1,2)} = \left| |\vec{G_1}| \pm |\vec{G_2}| \right|$. To relate the angle between consecutive crystals, we define $\varphi_{nm} \equiv \varphi_{n} - \varphi_{m}$. The intensity splitting distribution between the refracted beams (derived from the phenomenological theory presented in Ref. \cite{ebs2013}) is 
\begin{eqnarray}
I_2^{(1)} = I_{\chi}\cos^2 \left( \Phi_{\chi_2^{(1)}} - \Phi_{\chi} \right) = I_{\chi}\cos^2 \left( \frac{\varphi_{21}}{2} \right), \label{eqint1} \\ 
I_2^{(2)} = I_{\chi}\cos^2 \left( \Phi_{\chi_2^{(2)}} - \Phi_{\chi} \right) = I_{\chi}\sin^2 \left( \frac{\varphi_{21}}{2} \right), \label{eqint2}
\end{eqnarray}
where $I_{\chi} = I_{\chi,CP} = \frac{P_0} {A}$ for a circularly polarized input beam, while $I_{\chi} = I_{\chi,LP}= \frac{2P_0}{A} \cos^2 \left( \Phi_{\chi}-\Phi_0 \right) = 2 I_{\chi,CP} \cos^2 \left( \frac{\chi + \varphi_{1}}{2} -\Phi_0 \right)$ for the linearly polarized case. $P_0$ is the input beam's power, which redistributes over the area $A=4 \pi w_0 R$ occupied by the CR ring. Note that if $R=0$, then $A=\pi w_0^2$, where $w_0$ is the waist radius of the focused input beam. 
The two refracted beams after the BC$_2$ are also CR-filtered beams, that are described by their transverse positions $\vec{r}_{2}^{(1,2)} = \vec{r}(\chi_{2}^{(1,2)})$ and by the set of parameters $[\chi_{2}^{(1,2)},\Phi_{\chi_{2}^{(1,2)}}]$. Multiple rings formation is obtained by considering the full range $\chi \in [0, 2 \pi)$. In this case, Eqs.~(\ref{eq6}), (\ref{eqint1}) and (\ref{eqint2}) define two concentric rings with common center at $\vec{r}_{2}$, radii $R_{2}^{(i)}$ and azimuthal intensities $I_{2}^{(i)}$:
\begin{eqnarray}
\vec{r}_{2} = \vec{r}_0 + \vec{G}_1 + \vec{G}_2, \\ \label{eq2xtalsC}
R_{2}^{(i)} = \left| |\vec{G_1}| \pm |\vec{G_2}| \right|~(i=1,2), \\ \label{eq2xtalsR}
I_{2,CP}^{(1)} = \frac{P_0}{A_2^{(1)}} \cos^2 \left( \frac{\varphi_{21}}{2} \right), \label{eqint2cp1}\\ 
I_{2,CP}^{(2)} = \frac{P_0}{A_2^{(2)}} \sin^2 \left( \frac{\varphi_{21}}{2} \right), \label{eqint2cp2}
\end{eqnarray}
being $A_2^{(i)}=4 \pi w_0 R_2^{(i)}$ if $R_2^{(i)} \neq 0$ and $A_2^{(i)}=\pi w_0^2$ if $R_2^{(i)} = 0$. In other words, the second crystal shifts the center of the ring pattern and splits the parental CR ring into two concentric ones, as observed experimentally for the cascade of two crystals, see Fig.~\ref{fig5}(a). For input beams with linear polarization, the azimuthal intensity of the light ring patterns are: 
\begin{eqnarray} 
I_{2,LP}^{(1)} = 2 I_{2,CP}^{(1)} \cos^2 \left( \frac{\chi + \varphi_{1}}{2} - \Phi_0 \right), 
\label{eqint2lp1_1}\\ 
I_{2,LP}^{(2)} = 2 I_{2,CP}^{(2)} \cos^2 \left( \frac{\chi + \varphi_{1}}{2} - \Phi_0 + \frac{\pi}{2} \right),
\label{eqint2lp1_2}
\end{eqnarray}
for $|\vec{G_2}| > R_{1}$ and 
\begin{eqnarray}
I_{2,LP}^{(1)} = 2 I_{2,CP}^{(1)} \cos^2 \left( \frac{\chi + \varphi_{1}}{2} - \Phi_0 \right), \label{eqint2lp2_1}\\  
I_{2,LP}^{(2)} = 2 I_{2,CP}^{(2)} \cos^2 \left( \frac{\chi + \varphi_{1}}{2} - \Phi_0 \right), \label{eqint2lp2_2}
\end{eqnarray}
for $|\vec{G_2}| < R_{1}$.
Experimental patterns and corresponding theoretical simulations of a cascade of two biaxial crystals for input beams with linear and circular polarizations are shown in Fig.~\ref{fig5}. 

The third biaxial crystal (BC$_3$), once added into the cascade, splits each CR-filtered beam into two CR filtered beams with parameters defined by Eqs.~(\ref{eq4}), (\ref{eq5}), (\ref{eqint1}) and (\ref{eqint2}) as previously described. Therefore, four CR filtered beams appear at the positions defined by Eq.~(\ref{eq6}) with parameters $\vec{r}_{3} = \vec{r}_0 + \vec{G}_1 + \vec{G}_2 + \vec{G}_3$ and $R_{3}^{(i)} = \left| R_{2}^{(1,2)} \pm |\vec{G_3}| \right|$ ($i=1,2,3,4$). For the full range $\chi \in [0, 2 \pi)$ the latter results predict four concentric rings as shown in Fig.~\ref{fig5}(c), centered at $\vec{r}_{3}$ and with radii $R_{i}^{(3)}$ and corresponding azimuthal intensities $I_{3}^{(i)}$ as follows
\begin{eqnarray}
\vec{r}_{3} = \vec{r}_0 + \vec{G}_1 + \vec{G}_2 + \vec{G}_3, \label{eq3xtalsC}\\ 
R_{3}^{(i)} = \left| \left| |\vec{G_1}| \pm |\vec{G_2}| \right| \pm |\vec{G_3}| \right|~(i=1,2,3,4), \label{eq3xtalsR}\\ 
I_3^{(1)} = \frac{P_0}{A_3^{(1)}} \cos^2 \left( \frac{\varphi_{21}}{2} \right) \cos^2 \left( \frac{\varphi_{21}}{2} \right), \label{eq3xtals1}\\ 
I_3^{(2)} = \frac{P_0}{A_3^{(2)}} \sin^2 \left( \frac{\varphi_{32}}{2} \right) \cos^2 \left( \frac{\varphi_{21}}{2} \right), \label{eq3xtals2}\\ 
I_3^{(3)} = \frac{P_0}{A_3^{(3)}} \cos^2 \left( \frac{\varphi_{32}}{2} \right) \sin^2 \left( \frac{\varphi_{21}}{2} \right), \label{eq3xtals3}\\ 
I_3^{(4)} = \frac{P_0}{A_3^{(4)}} \sin^2 \left( \frac{\varphi_{32}}{2} \right) \sin^2 \left( \frac{\varphi_{21}}{2} \right), \label{eq3xtals4}
\end{eqnarray}
for CP input beams. For LP input beams, corresponding expressions can be obtained using Eqs.~(\ref{eqint2lp1_1})-(\ref{eqint2lp2_2}) and taking into account whether $|\vec{G}_{3}| < R_{2}^{(i)}$ or $|\vec{G}_{3}| > R_{2}^{(i)}$ ($i=1,2$). Explicit formulation is not presented here since for a cascade of $N=3$ BCs there are $3!=6$ possible combinations $\times$ 4 light rings $=24$ formulae. Experimental patterns and corresponding theoretical simulations for a cascade of three biaxial crystals for linearly and circularly polarized input beams are presented in Fig.~\ref{fig5}. 

Patterns for a cascade of $N$ biaxial crystals with characteristic vectors $\vec{G}_{i}$ can be obtained by applying Eq.~(\ref{eq3xtalsC}) (position of the center) Eq.~(\ref{eq3xtalsR}) (radii of each ring) and Eqs.~(\ref{eq3xtals1})-(\ref{eq3xtals4}) (intensity of each ring) recursively. In this case, up to $2^{N-1}$ concentric rings appear at the focal plane, as it was recently shown if Refs. \cite{phe2012,abd2011,Berry3} for circularly polarized input beams. We have checked this simple theory with cascades of up to five crystals.

In conclusion, pattern formation of multiple rings in CR can be described using a formalism based on double refraction of CR-filtered beams. We have confirmed that a cascade formed by $N$ BCs generates up to $2^{N-1}$ rings and we have presented a simple formulation that allows deducing the position, radii and intensity for each light ring. For the first time to our knowledge, analysis of the transverse intensity pattern of cascaded CR from a linearly polarized input beam has been reported, showing that the final pattern depends on the relative position of the BCs in the cascade, i.e, it is a non-commutable operation. The presented formulation can be of practical usefulness to easily predict the intensity patterns for light beams propagating through an arbitrary number of BCs.
 
The authors gratefully acknowledge financial support through Spanish MICINN contracts FIS2010-10004-E and FIS2011-23719, and the Catalan Government contract SGR2009-00347. A. T. acknowledges financial support through grant AP2010-2310 from the MICINN.

\end{document}